\documentclass[aps,prl,twocolumn,epsfig,groupedaddress]{revtex4}

\usepackage{epsfig}
\usepackage{amssymb,amsmath}
\begin{document}




\newcommand{\qs}{Q_{\rm sat}}
\newcommand{\qsa}{Q_{\rm sat, A}}

\title{Jet Shapes in Opaque Media}
\author{Antonio D. Polosa$^{(a)}$ and Carlos A. Salgado$^{(b)}$\footnote{Permanent address: Departamento de F\'\i sica de Part\'\i culas, Universidade de Santiago de Compostela (Spain).}}

\affiliation{$^a$INFN, Sezione di Roma, Roma, Italy \\
$^b$Dip. di Fisica, Universit\`a  di Roma ``La Sapienza'' and INFN, 
Roma, Italy\\}
\date{\today}

\begin{abstract}
We present general arguments, based on medium-induced radiative energy loss, which reproduce the non-gaussian shapes of away-side di-jet azimuthal 
correlations found in nucleus-nucleus collisions at RHIC. 
A rather simple generalization of the Sudakov form factors to opaque media allowing an effective description of the experimental data is proposed.
\end{abstract}
\maketitle


{\it Jet quenching}, the strong modification of the spectrum of particles produced at high transverse momenta in nucleus-nucleus collisions, 
is a solid experimental result of the RHIC program \cite{Adams:2005dq}. 
At present its main
observable is the strong suppression of the inclusive particle yield in the whole available range of $p_\perp\lesssim$ 20 GeV/c \cite{Adams:2005dq}\cite{Shimomura:2005en}. This suppression can be 
understood, in general terms, as due to the energy loss of highly energetic partons traversing the medium created in the collision. To further unravel the dynamical mechanism underlying this effect the questions of where and how the energy ``is lost" are of special relevance. The most promising experimental probes to investigate these issues are clearly related to the modification of the jet structures \cite{Salgado:2003rv}. 
In the most successful approach to explain jet quenching, the degradation of the energy of the parent parton produced in a elementary hard process, is due to the medium-induced  radiation of soft gluons \cite{Baier:1996sk}-\cite{Salgado:2003gb}, and  the energy transfer to the medium (recoil) is neglected. As usual, the medium-modified parton-shower will eventually convert into a hadron jet.
In the opposite limit, one could assume that a large fraction of the original parton energy is transferred to the medium, with a fast local equilibration, and diffused through sound and/or dispersive modes. The latter possibility has been advocated \cite{conical} as the origin of the striking non-gaussian shape of the azimuthal distributions in the opposite direction to the trigger particle \cite{Adler:2005ee}. 
In this paper we show that the same perturbative mechanism able to describe the inclusive suppression data, namely, the so-called radiative energy loss, can account for the experimentally observed two-peak shapes in the azimuthal correlations. The central point of our argument is the need of a more exclusive treatment of the distributions to describe experimental data with restrictive kinematical constrains. To this end, we supplement the medium-modified jet formation formalism with the Sudakov suppression form factor following the well known in-vacuum approach. Furthermore, we assume that the final hadronic distributions follow the parton level ones (parton-hadron duality).

{\it The medium-induced gluon radiation}. 
Let us suppose that a very
energetic parton propagating through a medium of length $L$ emits a gluon
with energy $\omega$ and transverse momentum $k_\perp$ with respect to the fast
parton. The formation time of the gluon is $t_{\rm form}\sim \frac{2\omega}{k_\perp^2}$. Its typical transverse momentum is 
\begin{equation}
k_\perp^2\sim \frac{\langle q_\perp^2\rangle_{\rm med}}{\lambda}\, t_{\rm form}\sim 
\sqrt{2\omega\hat q}.
\label{eqkt2}
\end{equation}
$\hat q$ is the transport coefficient \cite{Baier:1996sk}
which characterizes the
medium-induced transverse momentum squared $\langle q_\perp^2\rangle_{\rm
med}$ transferred to the projectile per unit path length $\lambda$. In this picture the gluon acquires transverse momentum because of the Brownian-motion in the transverse plane due to multiple soft scatterings during $t_{\rm form}$.  The typical emission angle 
\begin{equation}
\sin\theta\equiv \frac{k_\perp}{\omega}\sim \left(\frac{2\hat q}{\omega^3}\right)^{1/4}
\label{eq:emangle}
\end{equation}
defines a minimum emission energy $\hat\omega\sim (2\hat q)^{1/3}$ below which the radiation is suppressed by formation time effects \cite{Salgado:2003gb}.  The latter is a crucial observation for the discussion to follow.  Notice that for energies smaller
than $\hat\omega$ the angular distribution of the medium-induced emitted gluons peaks at large values. In \cite{Eskola:2004cr} 
a fit to the inclusive high-$p_\perp$ particle suppression gives the values
$\hat q\sim 5...15$ GeV$^2$/fm for the most central AuAu collisions at RHIC, i.e.,   $\hat\omega\sim$ 3 GeV [we have taken $\hat\omega\sim 2(2\hat q)^{1/3}$ as indicated by numerical results]. This situation can be computed from the spectrum of the medium-induced gluon radiation in the limit $k_\perp^2< \sqrt{2\omega\hat q}$, $\omega\ll\omega_c\equiv\frac{1}{2}\hat q L^2$. This gives in the multiple soft scattering approximation \cite{Wiedemann:2000za,Salgado:2003gb}
\begin{equation}
\frac{dI^{\rm med}}{d\omega\,dk^2_\perp}\simeq\frac{\alpha_s C_R}{16\pi}\,L\,\frac{1}{\omega^2}.
\label{eq:cohspec}
\end{equation}
Compared to the corresponding spectrum in the vacuum, the one in
(\ref{eq:cohspec}) is softer and the typical emission angles larger, producing a broadening of the jet signals. 

{\it The parton shower evolution}. Equation (\ref{eq:cohspec}) gives the inclusive spectrum of gluons emitted by a high-energy parton traversing a medium. For practical applications, however, more exclusive distributions, giving the probability of one, two... emissions are needed. How to construct such probabilities, using Sudakov form factors, is a well known procedure in the vacuum. In the medium, a first attempt to deal with the strong trigger bias effects in inclusive particle production, was that of using an independent gluon emission approximation with corresponding Poissonian probabilities \cite{Baier:2001yt}. Here we propose to improve this assumption by including the virtuality evolution through medium--modified Sudakov form factors. It is encouraging to notice that the original formulation is recovered once the virtuality evolution is neglected. 

Since we are interested in angular distributions the parton shower description proposed in \cite{Marchesini:1983bm} is particularly convenient. Let us introduce the evolution variable 
\begin{equation}
\xi=\frac{q_1\cdot q_2}{\omega_1\omega_2}
\end{equation}
to define the branching of a particle (gluon in general) with 
virtuality $q$ and energy $\omega$ in two particles 
of virtuality $q_1,\ q_2<q$ and energies $\omega_1\equiv
z\omega$, $\omega_2\equiv (1-z)\omega$. If $\omega_1,\ \omega_2
\gg m_1,\ m_2$, as we will assume, $\xi\simeq 1-\cos\theta_{12}$, with $\theta_{12}=\theta_1+\theta_2$.  $\theta_1$, $\theta_2$ are the angles formed
by the daughter partons with the parent parton.
Consequently
\begin{equation}
\frac{dq^2}{q^2}=\frac{d\xi}{\xi},
\label{eq:jaco}
\end{equation}
i.e.,  the virtuality evolution can be converted into an evolution in the variable $\xi$.
The corresponding probability distribution for one branching is
\begin{equation}
d{\cal P}(\xi,z)=\frac{d\xi}{\xi}\,dz\,\frac{\alpha_s}{2\pi}P(z)\,
\Delta(\xi_{\rm max},\xi)\theta(\xi_{\rm max}-\xi)\theta(\xi-\xi_{\rm min})
\label{eq:sudakxi}
\end{equation}
where $P(z)$ is the splitting function. The Sudakov form factor 
controlling the evolution is~\cite{bks}:
\begin{equation}
\Delta(\xi,\omega)=\exp\left\{-\int_\xi^{\xi_{\rm max}}\frac{d\xi^\prime}{\xi^\prime}
\int_\epsilon^{1-\epsilon} dz\,\frac{\alpha_s}{2\pi}\,P(z)\right\}
\label{eq:sudakreg}
\end{equation}
and $\epsilon=Q_0/\omega\sqrt{\xi}$ with $Q_0$ a cut-off. $\Delta(\xi,\omega)$ can be interpreted as the probability of no branching between the scales $\xi$ and $\xi_{\rm max}$. 

We propose to generalize the above formulation to the in-medium case by noticing that in our notation \cite{Wiedemann:2000za}\cite{Salgado:2003gb}
\begin{equation}
\frac{dI^{\rm vac}}{dz dk_\perp^2}=\frac{\alpha_s}{2\pi}\frac{1}{k_\perp^2}P(z).
\label{eq:specvac}
\end{equation}
We define the corresponding equations (\ref{eq:sudakxi}), (\ref{eq:sudakreg}) for the medium by changing $dI^{\rm vac}/dzdk^2_\perp$ to $dI^{\rm med}/dzdk^2_\perp$ given by Eq.~(\ref{eq:cohspec}). We have clear motivations to 
make this {\it Ansatz}: (i) it provides a clear probabilistic interpretation with the right limit to the most used ``quenching weights'' \cite{Baier:2001yt}\cite{Salgado:2003gb} when the virtuality is ignored; (ii) the evolution equations in the case of nuclear fragmentation functions in DIS can be written as usual DGLAP equations with medium-modified splitting functions $P(z)\to P(z)+\Delta P(z)$ \cite{Wang:2001if}. In our case Eq.~(\ref{eq:cohspec}) would correspond to $\Delta P(z)$ with the appropriate factors.

{\it Results}. In what follows, we present some simple analytical estimates
following the approach just sketched.
The general case of arbitrary $\omega_1$ and $\omega_2$ is difficult to study analytically. 
Let us study, instead, two extreme cases: (1) one of the particles takes most of the incoming energy $\omega_1\gg\omega_2$ and, hence, $\theta_1\simeq 0$ -- we name this as ``J-configuration''; (2) the two particles share equally the available energy, $\omega_1\simeq\omega_2$ and $\theta_1\simeq\theta_2$ -- named as 
``Y-configuration''.

Let us first consider the case (1) where $\xi=1-\cos\theta_{12}\equiv 1-\cos\theta$ with $\sin\theta=k_\perp/\omega$. From (\ref{eq:cohspec}) we get
\begin{equation}
\frac{dI}{dzd\xi}=\frac{\alpha_s C_R}{8\pi}\, E\,L\,(1-\xi).
\label{eq:case1}
\end{equation}
%
Therefore  the corresponding Sudakov form factor reads:
\begin{equation}
\Delta^{\rm med}(\xi_{\rm max},\xi)=\exp\left\{-\frac{\alpha_s C_R}{8\pi}
\,L\,E\int_\xi^{\xi_{\rm max}}
d\xi^\prime\,(1-\xi^\prime)\right\}
\end{equation}
where the small contribution form the integration in $z$ has been neglected [we have checked that the results does not vary as long as $\omega\gg Q_0$]; $\alpha_s=1/3$ will be assumed in the following. 
Taking $\xi=1-\cos\theta$, $\xi_{\rm max}=1$ and inserting into the probability of splitting, Eq.~(\ref{eq:sudakxi}), we get:
\begin{equation}
\frac{d{\cal P}(\theta,z)}{dz\,d\theta}=\frac{\alpha_s C_R}{8\pi}\,E\,L\,\sin\theta\cos\theta
\exp\left\{-E\,L\,\frac{\alpha_s C_R}{16\pi}\cos^2\theta\right\}.
\label{eq:splitms}
\end{equation}
Eq.~(\ref{eq:splitms}) is the probability distribution
for a parent parton to split just once, emitting a gluon at angle $\theta$ 
with fraction $z$ of the incoming momentum.

\begin{figure}
\begin{center}
\includegraphics[width=0.3\textwidth,angle=-90]{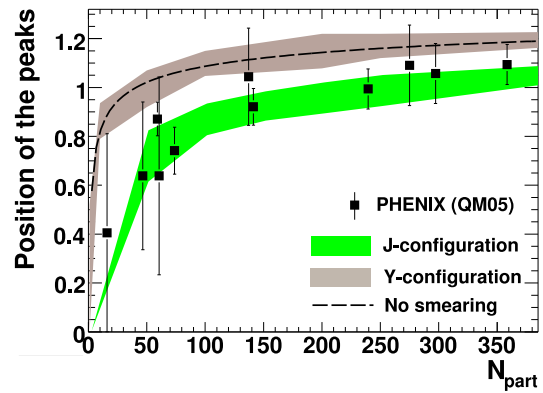}
\end{center}
\caption{Position of the peaks of the $\Delta\Phi$-distribution and comparison with PHENIX data from Ref. \cite{Grau:2005sm} for the J- and Y-configurations. 
We also show the analytical estimate (\ref{eq:phimax}) without any smearing.}
\label{fig:dpar1}
\end{figure}

The variables of (\ref{eq:splitms}) are polar coordinates with respect to the jet axis (the direction of the parent parton). Assuming that this parent parton was produced at 90$^o$ in the center of mass frame of the collision, we can transform the coordinates of the emitted gluon to the laboratory polar coordinates $\Phi$ and $\theta_{\rm lab}$ with respect to the beam direction $\hat{z}$. Normally, one uses the pseudorapidity $\eta=-{\rm log\,tan\,}(\theta_{\rm lab}/2)$. The jabobian is
\begin{equation}
d\theta d\beta=\frac{d\eta d\Phi}{{\rm cosh}\eta\sqrt{{\rm cosh}^2\eta-\cos^2\Phi}}
\end{equation}
where $\beta$ was integrated in the previous expressions to give $2\pi$ in the spectra. Taking, for simplicity $\eta=0$, in the most favorable detection region, the answer is simply
\begin{equation}
\frac{d{\cal P}(\Phi,z)}{dz\,d\Phi}\Bigg\vert_{\eta=0}=
\frac{\alpha_s C_R}{16\pi^2}\,E\,L\,\cos\Phi
\exp\left\{-E\,L\,\frac{\alpha_s C_R}{16\pi}\cos^2\Phi\right\}
\label{eq:splitlab}
\end{equation}
giving the probability of one splitting as a function of $\Phi$.  Thus we reach our objective: the possibility of describing non-trivial angular dependences, as shown in Eqs.(\ref{eq:splitms}) and (\ref{eq:splitlab}) with a perturbative mechanisms. The distribution found has two maxima whose positions are determined by:
\begin{equation}
\Phi_{\rm max}=\pm {\rm arccos}\sqrt{\frac{8\pi}{E\,L\,\alpha_s\, C_R}}
\label{eq:phimax}
\end{equation}
\begin{figure}
\includegraphics[width=0.3\textwidth,angle=-90]{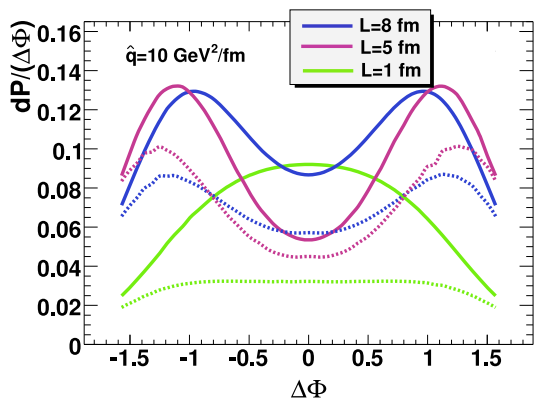}
\caption{The probability of just one splitting (\protect\ref{eq:splitms}) as a function of the laboratory azimuthal angle $\Delta\Phi$ for a gluon jet of $E_{\rm  jet}=7$ GeV. Different medium lengths are plotted for the  J- and Y-configurations (solid and dotted lines respectively).}
\label{fig:ndist}
\end{figure}
The angular shape found in (\ref{eq:splitlab}) is very similar to the one found experimentally. We do not intend to perform a  detailed calculation of the experimental situation. We just improve our calculation by introducing a  simple model to take into account the additional smearing of the jet shape introduced by the triggering conditions. Setting $\eta=0$ for the trigger particle, we take into account (i) the uncertainty due to the boosted center of mass frame of the partonic collision by integrating the probability (\ref{eq:splitlab}) in $2\Delta\eta$ with $\Delta\eta=1$ \protect\footnote{We have checked that a convolution with a gaussian as given in \protect\cite{Renk:2006mv} does not affect our results.}; (ii) an additional uncertainty in azimuthal angle $\Delta\Phi$ given by a gaussian with $\sigma$= 0.4 \cite{Adler:2006sc}. Specifically, we take
\begin{equation}
\frac{dP}{d\Delta\Phi dz}=\frac{1}{N}\int_{-\Delta\eta}^{\Delta\eta}d\eta\int
d\Phi'\frac{d{\cal P}}{d\Phi' dz d\eta}{\rm e}^{-\frac{(\Delta\Phi-\Phi')^2}{2\sigma^2}}
\end{equation}
$N=2\Delta\eta\sqrt{2\pi\sigma^2}$ being a normalization factor.
The results are plotted in Fig. \ref{fig:dpar1} for three different medium lengths and $E_{\rm jet}=7$ GeV [the quoted value $\hat q=10$ GeV$^2$/fm assures that the results hold for gluon energies $\omega<\hat\omega\simeq 3$ GeV]. In order to compare with the centrality dependence of the position of the maxima measured experimentally \cite{Grau:2005sm} we simply take $L=N_{\rm part}^{1/3}$. Although this geometric procedure is too simplistic, it serves us to have a feeling of the description of the data plotted in Fig. \ref{fig:dpar1}. We must also point out that the centrality dependence of the transport coefficient is not taken into account in these figures. The reduction of $\hat q$ with centrality ($\hat{q}\sim dN/dy\sim N_{\rm part}$) makes the radiation more collinear when $\hat q^{1/3}<\omega$, i.e., the radiation will be more collinear than plotted for decreasing centrality at fixed $\omega\simeq p_\perp^{\rm assoc}$.

Let us now consider the case of the Y-configuration, where $\omega_1\simeq\omega_2$ and the two angles are similar $\theta_1\simeq\theta_2\equiv\theta$. Now the variable
$\xi=1-\cos\theta_{12}=1-\cos 2\theta$ and $k_\perp^2/\omega^2=\sin^2\theta=\xi/2$. Changing variables we get
\begin{equation}
\frac{dI}{dzd\xi}=\frac{\alpha_s C_R}{32\pi}\,E\,L .
\end{equation}
Repeating the same procedure followed before
\begin{equation}
\frac{d{\cal P}(\Phi,z)}{dzd\Phi}=\frac{C_R\alpha_s}{64\pi^2}\,E\,L\,\exp\left\{-\frac{C_R\alpha_s}{32\pi}\,E\,L\cos\Phi\right\}.
\end{equation}
In this case, the maxima are outside the borders of the physical phase space but still a minimum at $\Phi=0$ occours. Anyway, the smeared distribution does present  maxima. The corresponding curves and positions of the maxima are plotted in Figs. \ref{fig:ndist} and \ref{fig:dpar1}. 

{\it The picture} which emerges from our analysis is the following: the splitting probability of a highly energetic parton produced inside the medium by a hard process presents well defined maxima in the laboratory azimuthal angle when the emitted gluons have energies $\omega<\hat\omega$, with $\hat\omega\sim $ 3 GeV for the most central collisions at RHIC. This is a reflection of the jet broadening predicted long ago as a consequence of the medium-induced gluon radiation \cite{Baier:1996sk}-\cite{Salgado:2003gb}. In the case that the experimental triggering conditions are such that only a small number of splittings are possible (by restrictive kinematical constrains, e.g. $p_\perp^{\rm assoc}\simeq p_\perp^{\rm trigg}$) these structures should be observed. Our estimate, taking into account the parton-hadron duality, is then $p_\perp^{\rm assoc}\lesssim \hat\omega\simeq$ 3 GeV. When, on the contrary, the allowed number of splittings is larger, we expect this angular structure to disappear, filling the dip located in correspondence of the jet axis. The reason is that the inclusive spectrum (which does not present a two-peak structure) should be recovered in these conditions. This could explain the absence of a two-peak structure when the kinematical cut on $p_\perp^{\rm assoc}$ is relaxed (or/and when $p_\perp^{\rm trigg}$ is larger for fixed $p_\perp^{\rm assoc}$). In this situation the broadening is, however, still present with a flatter distribution which extends in a large range of $\Delta\Phi$. A realistic comparison with experimental data would need a much more sophisticated analysis, including not only the probability of multiple splittings but also the hadronization here ignored.
Our results are very encouraging though as the distributions we obtain for the single splitting resemble very much the experimental findings, both in shape and in centrality dependence. 

Let us also comment on the relation with other approaches. In \cite{Vitev:2005yg} the large angle medium-induced gluon radiation has been considered and modifications of the angular distribution computed. The obtained jet structures are, however, basically gaussian in azimuthal angle, the effect of the medium being a broadening on the typical jet width. Two main differences with our approach are the source of this discrepancy: First, and most importantly, the angular distributions
 in \cite{Vitev:2005yg} are given by the inclusive spectrum, which does not present the two-peak structure in $\Delta\Phi$ --
the main point of the present paper is to provide a proposal on how to improve this description by the inclusion of Sudakov form factors; Second, in the single hard approximation used in \cite{Vitev:2005yg} for the medium-induced gluon radiation the typical emission angle {\it decreases} with increasing in-medium path-length \cite{Majumder:2005sw} as $\sin\theta\sim1/\sqrt{L}$, making the radiation more and more collinear with increasing centrality (in contrast with (\ref{eq:emangle}) which is independent on the centrality). 

On the other hand, one of the most popular explanations for the non-gaussian shape of the away-side jet-signals found at RHIC is in terms of shock waves produced by the highly energetic particle into the medium. In this picture, a large amount of the energy lost must be transferred to the medium which thermalizes almost instantaneously. In general the released energy excites both sound and dispersive modes, and only the first ones produce the desired cone-signal (in the dispersive mode, the energy travels basically collinear with the jet). The energy deposition needed for the sound modes to become visible in the spectrum has been found to be quite large \cite{Casalderrey-Solana:2006sq}. To our knowledge, no attempt has been made so far to describe the centrality dependence of the shape of the azimuthal correlations in this approach. 
Given the fact that the two formalisms described above rely on completely different hypotheses, 
finding experimental observables which could distinguish between them is certainly an issue which deserves further investigation. New data on three particle correlations are expected to shed some light on the problem. Here, we just notice that our Y-configurations would lead to similar signatures as the ones from the shock wave model. In the most general case, however, different configurations, not considered in our simple analysis, and characterized by different radiation angles for both gluons, would produce a smeared signal.

CAS is supported by the 6th Framework Programme of the European Community under the Marie Curie contract MEIF-CT-2005-024624.

\end{document}